\documentclass{aa}  

\usepackage{graphicx}
\usepackage{txfonts}
\usepackage{lipsum}
\usepackage{subcaption}         % necessary for continued figures, example in section 3
\usepackage{lscape}             % to rotate a single page table, example in appendix.
\usepackage{placeins}           % useful with \FloatBarrier, to keep 
\usepackage[colorlinks,citecolor=blue,urlcolor=blue,filecolor=blue,linkcolor=blue]{hyperref}
\usepackage{txfonts}
\usepackage{siunitx}

\addto\extrasenglish{
    
}
\addto\extrasenglish{
    
}
\addto\extrasenglish{
    
} 
\addto\extrasenglish{
    
}

\begin{document}

\title{Multi-wavelength insights into the pulsar wind nebula candidate near 1LHAASO J0343+5254u: an obscured merging galaxy cluster?}
\titlerunning{Insights into the PWN candidate near 1LHAASO J0343+5254u: an obscured merging galaxy cluster?}

   \author{H. W. Edler\inst{1}
        \and M. Arias\inst{1}
        \and A. Botteon\inst{2}
        \and C. G. Bassa\inst{1}
        }
       \institute{ASTRON, Netherlands Institute for Radio Astronomy, Oude
    Hoogeveensedijk 4, 7991 PD Dwingeloo, The Netherlands\\
                  \email{edler@astron.nl}
                              \and
                                INAF - Istituto di Radioastronomia,
              via P. Gobetti 101, Bologna, Italy}

   \date{Received December 18, 2025; accepted March 2, 2026}

  \abstract
   {The advent of the Large High Altitude Air Shower Observatory (LHAASO) accelerated the detection of TeV and PeV gamma-ray sources. Some of these are associated with pulsar wind nebulae (PWNe) and other Galactic objects, while others are yet to be connected to sources at other wavelengths. Recently, the discovery of an extended X-ray source within the unidentified PeV source 1LHAASO~J0343+5254u was reported, this source was claimed as a candidate PWN based on its X-ray spectrum.}
   {We will revisit the interpretation of the extended X-ray source based on  multi-wavelength observations.}
   {We present new LOFAR continuum radio imaging at observing frequencies of 54 and 144\,MHz, an alternative X-ray modeling and archival near-infrared (NIR) data.}
   {We discover several radio sources with morphologies and spectra suggestive of a radio halo, a radio relic and tailed radio galaxies, all of which are typically found in (merging) galaxy clusters. Furthermore, we show that the X-ray data can be modeled as thermal emission from the intracluster medium (ICM), with our best-fitting thermal ICM model being slightly preferred to a non-thermal power-law fit. We further find a $9.7\sigma$ over-density in red NIR sources in the surrounding region, among them possible hosts of the tailed radio sources. }
   {{Our results favor an interpretation of the X-ray source as a massive, merging galaxy cluster located in a highly extinct region of the Galactic plane, unrelated to 1LHAASO~J0343+5254u. Future observations in the hard X-ray regime will be able to conclusively settle the discussion on the nature of the X-ray emission.}}

   \maketitle

\nolinenumbers

\section{Introduction}

Ultra High Energy (UHE; $E>100$~TeV) gamma-rays, a long inaccessible window of the electromagnetic spectrum, are now routinely observed thanks to ground-based extensive air shower arrays,
most notably the Large High Altitude Air Shower Observatory \cite[LHAASO,][]{aharonian21}. The recently published First LHAASO Catalog of Gamma-ray Sources \citep{cao24} lists 90 sources with $E>0.1$~TeV, of which 43 show UHE emission.
The LHAASO results have called into question the long-held idea 
that supernova remnants (SNRs) were the main accelerators of cosmic rays (CRs) up to energies of a few PeVs \citep{hillas05}. Despite many instances of $\leq\,$TeV gamma-ray sources being associated with SNRs, other sources, such as pulsar wind nebulae \cite[PWNe,][]{abeysekara20,cao21}, pulsar TeV halos \citep{linden17,lopez-coto22}, and superbubbles \citep{abeysekara21} are now also contemplated as likely Galactic PeVatrons\footnote{PeVatrons accelerate cosmic rays to energies beyond 1\,PeV and are traced by UHE emission.}, contributing non-negligible amounts to the total Galactic CR flux. There is strong community interest in understanding the nature of the  LHAASO sources; this requires a multi-wavelength approach.

In this context, \citet[hereafter DK25]{dikerby25} observed a region in the Galactic plane towards ${l{=}146.94^\circ,\,b{=}{-}1.63^\circ}$ in the X-rays with XMM-Newton. This region contains two sources cataloged in \cite{cao24}: 1LHAASO~J0339+5307 and 1LHAASO~J0343+5254u, originally published as the single source LHAASO~J0341+5258 in \citet{Cao2021b}. The latter of these two sources also shows UHE emission.
As a result of their observing campaign, DK25 discovered the new X-ray source XMMU\,J034124.2+525720, centered at $\alpha_\mathrm{J2000}{=}03^\mathrm{h}41^\mathrm{m}24^\mathrm{s}$, $\delta_\mathrm{J2000}{=}{+}52\degr57\arcmin10\arcsec$  and positionally coincident with the lower-energy component of 1LHAASO~J0343+5254u. 
This source is extended, and its spectrum was fitted with a power-law model, suggesting that it may be a PWN. DK25 found no clear X-ray pulsations or radio counterpart, and no Fermi gamma-ray source at the source location. Based on the X-ray spectrum and morphology, they classified the X-ray source as a PWN candidate and a possible counterpart of the UHE source.

An alternative source of extended X-ray emission, not associated with UHE photons, is diffuse thermal emission from the ICM of a galaxy cluster. Deep within the Galactic plane, the strong absorption of soft X-rays ($E < 2\,$keV) can cause the spectra of PWNe and galaxy clusters to resemble each other.
This scenario was not explored by DK25, but considered as an alternative to a PWN origin by earlier studies of similar sources \citep{Katsuda2012DiscoveryPulsarWind,Halpern2019NoPulsarWind}. 
Radio observations provide an independent method for identifying highly obscured clusters, as they are not affected by Galactic extinction \citep{Kollatschny2021}.
Radio emission in clusters can originate from cluster member galaxies. Tailed radio galaxies (TRG), which are shaped by their interaction with the ICM, are a typical phenomenon observed in clusters. Clusters can further show diffuse radio emission associated with merger activity \citep{Cuciti2021}. Depending on the morphology and location, diffuse cluster radio sources are classified as radio relics (elongated sources in the periphery) or radio halos \cite[extended sources co-spatial to the X-ray emission,][]{vanWeeren2019DiffuseRadioEmission}. These sources have steep spectra (spectral indices $\alpha < -1$\footnote{We follow the $S\propto\nu^\alpha$ convention.}).

On the other hand, the synchrotron emission of PWNe results from accelerated particles injected by its central pulsar. The emission mechanism is the same across many orders of magnitude, typically resulting in extended sources that shrink in size with increasing energy due to the effect of synchrotron losses. The morphology of the extended source can vary, showing shapes that are circular, elliptical, tailed, or irregular, but the emission is typically contiguous and overlaps with the location of the pulsar, which injects the emitting particles.
The observed radio spectral indices for PWNe range between $-0.3 \lesssim \alpha \lesssim 0$ \citep{gaensler06}.

In this study, we will revisit the classification of the X-ray source discovered by DK25 based on additional multi-frequency data.  We assume a flat $\mathrm{{\Lambda}CDM}$ cosmology with $\Omega_\mathrm{m}=0.3$ and $H_0=70\,\mathrm{km\,s^{-1}\,Mpc^{-1}}$. This work is organized as follows: in \autoref{sec:analysis}, we present our analysis of the X-ray, radio and NIR data. We discuss our results in \autoref{sec:discussion} and conclude in \autoref{sec:conclusion} .
\section{Observations and analysis}\label{sec:analysis}
\subsection{X-ray analysis}\label{sec:xray}
The region surrounding the 1LHAASO J0343+5254u was observed by XMM-Newton (IDs: 0923400401, 0923400801, and 0923401401) for ${\sim}115\,$ks. These observations were originally presented in DK25.
We reprocessed the data following the procedure of \cite{bartalucci23}, using the Extended Source Analysis Software \cite[ESAS;][]{snowden08} within the XMM-Newton Scientific Analysis System (SAS v16.1). After filtering the observations for soft proton flares, we produced exposure-corrected EPIC images for each ObsID in the $0.5-7.0$\,keV band and combined them into a single mosaic. We note that XMMU\,J034124.2+525720 lies $>$10 arcmin from the center of the observation, where the PSF of the instrument degrades. 

DK25 modeled the X-ray spectrum of the extended emission with a power-law. Here, we investigate whether the data can instead be described as thermal emission from the ICM, using the same spectral extraction regions used in DK25 for both the source and the background. We first modeled the background following Sec. 4 of \citet{Rossetti2024}. As DK25, we found that the Galactic $N_{\rm H}$ value reported by \citet{HI4PICollaboration2016HI4PIFullskySurvey} does not yield a satisfactory fit;  therefore, we allowed it to vary during the background modeling. Our best-fit value, which we adopt hereafter, is $N_{\rm H} = 1.58\times10^{22} {\rm\,cm^{-2}}$, consistent within uncertainties with that found by DK25.

\begin{figure}
    \centering
    \vspace{2mm}
    \includegraphics[width=0.75\linewidth,angle=-90]{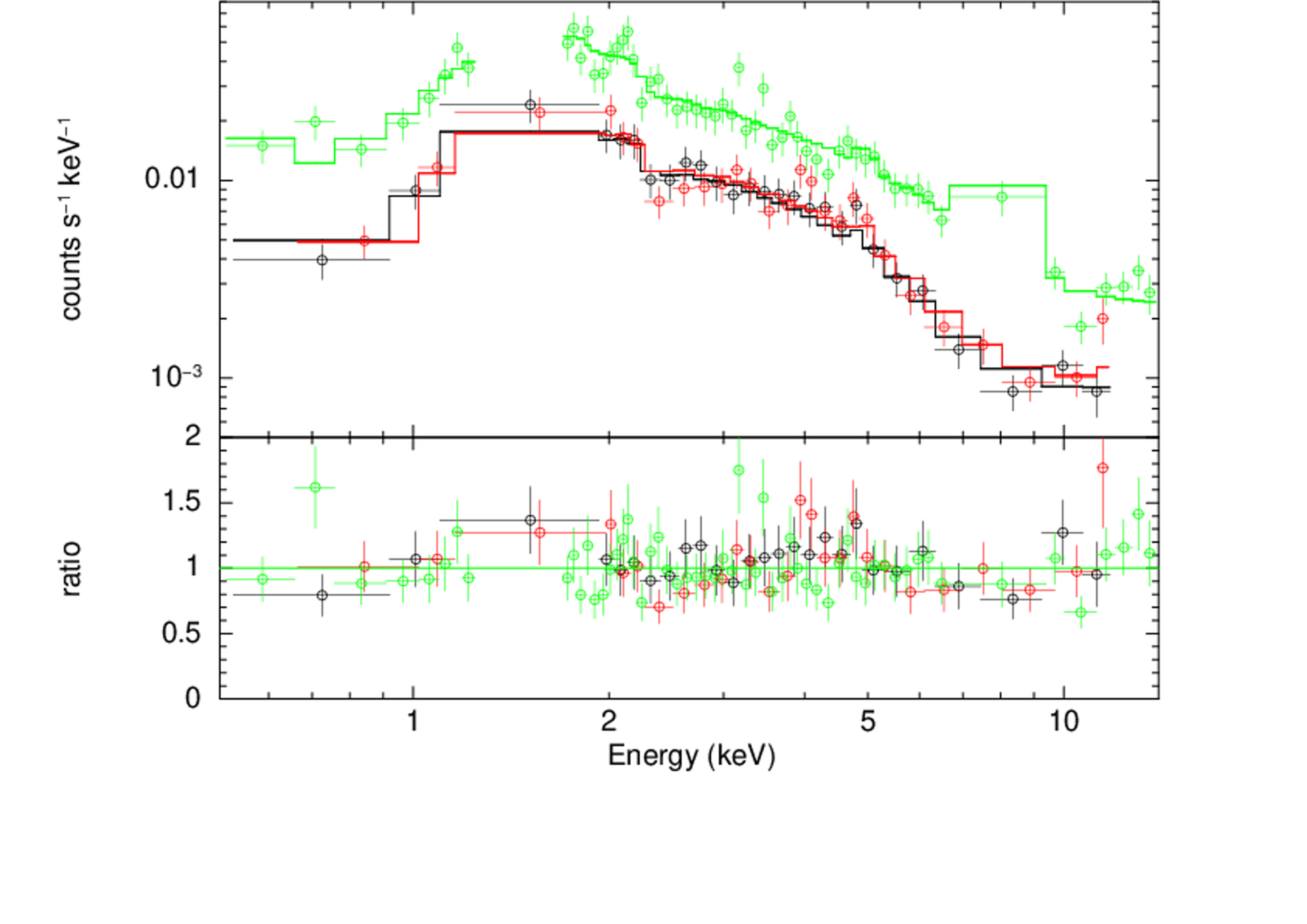}
    \vspace{-8mm}
    \caption{Observed source spectrum (black=MOS1, red=MOS2, green=pn) with APEC best-fit model reported. The spectra of the different cameras and observations were fitted jointly, but the results of a single ObsID (0923400401) were reported for clarity. For energy ranges 1.2-1.9 keV (for the MOS detectors) and 1.2-1.7 keV and 7-9.2 keV (for the pn detector) were excluded during the fit. Each spectral bin was rebinned to a minimum significance of 5$\sigma$ for visualization purposes. Residuals in the pn spectrum at 2.1\,keV are instrumental \citep[Au lines,][]{2008A&A...486..359L}.}
    \label{fig:xrayspec}
\end{figure}

With the background model fixed, we proceeded to fit the spectrum of the primary source. Using a power-law model, we obtained a photon index $\Gamma = 1.75 \pm 0.03$, in agreement with DK25. As alternative, we consider that the X-ray source is thermal emission from the ICM, represented by the APEC model \citep{smith2001}. To obtain a robust fit given the data quality, we limit the number of free parameters of the model and assume a typical metal abundance of $Z = 0.3 Z_\odot$ \citep[e.g.][]{Molendi2016}.
The best-fitting APEC model has a temperature of $kT = 8.51^{+0.75}_{-0.47}$ keV and a redshift of $z = 0.371^{+0.010}_{-0.006}$. The resulting statistics are C-stat/degrees of freedom = 1817.68/1798 for the power-law  (shown in \autoref{fig:xrayspec_power}) and 1809.66/1797 for the APEC fit (shown in \autoref{fig:xrayspec}). Both models provide an acceptable description of the data, with the thermal model being slightly preferred (at a moderate significance of 2.8$\sigma$ based on the likelihood ratio test).
We also repeated the APEC fit leaving the metallicity unconstrained. The resulting best-fitting values are a temperature of $kT = 8.6^{+0.7}_{-0.5}$ keV, a redshift of $z = 0.371^{+0.010}_{-0.008}$ and a metallicity of $Z = 0.23^{+0.07}_{-0.06}$ (C-stat./d.o.f. 1808.85/1796), within uncertainties of the values found for the constrained fit.  
\subsection{LOFAR imaging}\label{sec:radio}
\begin{figure*}
        \centering
    \includegraphics[width=1.0\linewidth]{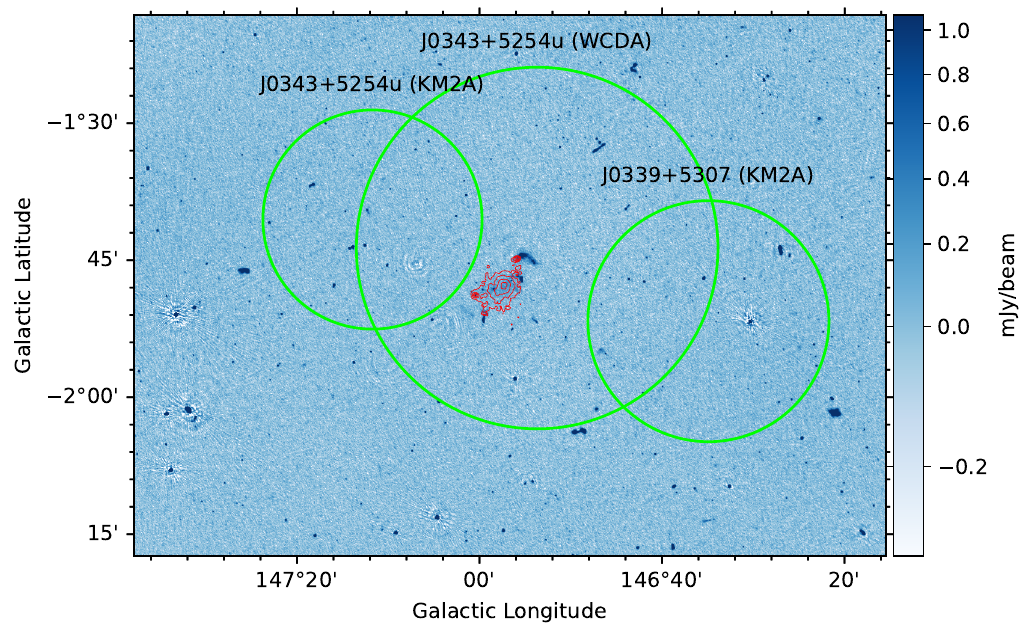}
    \caption{LoTSS DR3 of the wider area surrounding the LHAASO sources in Galactic coordinates. The central frequency is 144\,MHz and the noise level is $\sigma_\mathrm{rms}=100\,\mathrm{\mu{Jy}\,beam^{-1}}$ at a resolution of $6''$. The green circles show the 39\% containment radius of the LHAASO sources, and the red contours mark the $0.5-7$\,keV emission around XMMU\,J034124.2+525720 smoothed to 20$''$ resolution, starting at $5\times10^{-6}\,\mathrm{cts\,s^{-1}\,arcsec^{-2}}$ and increasing in factors of two.}
    \label{fig:lotss}
\end{figure*}
\begin{figure*}
\sidecaption
  \includegraphics[width=12cm]{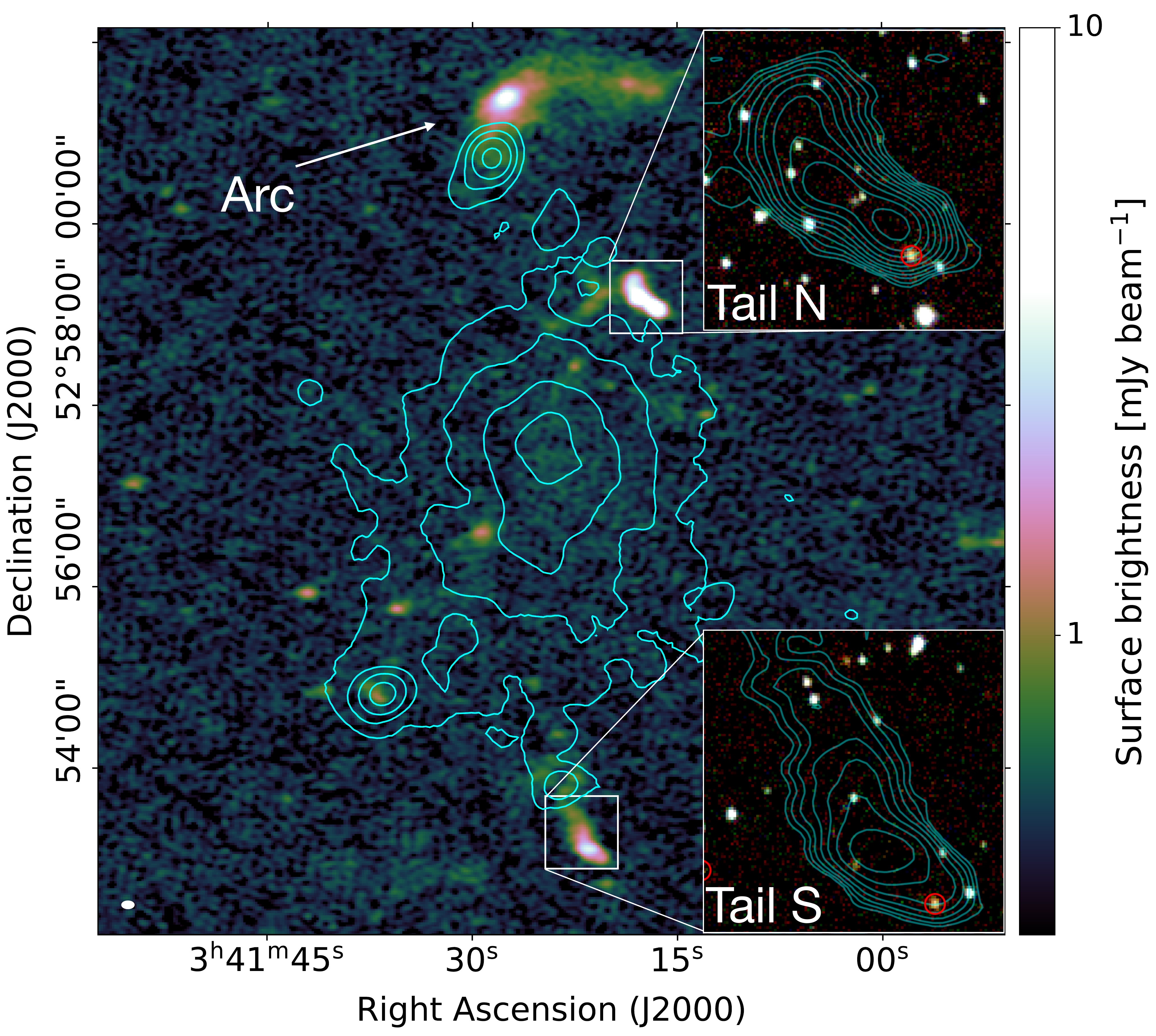}
    \caption{Radio map at 144\,MHz, the noise level is $\sigma_\mathrm{rms}=100\,\mathrm{\mu{Jy}\,beam^{-1}}$ at $5''\times8''$ resolution. In cyan we show the $0.5-7$\,keV X-ray contours. The insets show the NIR image with the radio contours of the tailed sources overlaid; red circles mark UGS galaxies with red $(J-K)$ color. }    \label{fig:combined}
\end{figure*}

\begin{table}[]
    \centering
    \caption{Flux densities and spectral indices.}
    \begin{tabular}{l c c c}
        Source & $S_{54}$ [Jy] & $S_{144}$ [Jy] & $\alpha$  \\\hline
        Arc & $0.393\pm0.041$ & $0.123\pm0.012$ & $-1.18\pm0.15$\\
        Tail N & $0.179\pm0.019$ & $0.069\pm0.007$ & $-0.97\pm0.15$\\
        Tail S & $0.082\pm0.011$ & $0.033\pm0.003$  & $-0.93\pm0.17$\\
        Diffuse halo & $0.084\pm0.018$ & $0.027\pm0.003$  & $-1.36\pm0.38$\\
    \end{tabular}
    \label{tab:fluxes}
\end{table}
The location of XMMU\,J034124.2+525720 is covered by the third data release of the LOFAR Two-metre Sky Survey (LoTSS; \citealt{Shimwell2026}) at 144\,MHz as well as in the upcoming data release of the LOFAR Low-band Antenna Sky Survey (LoLSS; \citealt{deGasperin2023LOFARLBASky}, de Gasperin+ in prep.) at 54\,MHz.
The LoTSS map of the wider region surrounding the XMM and LHAASO sources is shown in \autoref{fig:lotss}. Extended radio sources are visible at the location of the X-ray source.
For further analysis, we re-calibrated the LoTSS and LoLSS observations covering these extended radio sources, details of the calibration are provided in \autoref{sec:app_radio}.
In \autoref{fig:combined}, we display the re-calibrated 144\,MHz map with the X-ray contours overlaid in cyan. A prominent, arc-like structure is visible towards the north, containing a region with increased surface brightness in the east. We also detect two tailed sources (white squares in the figure). 
Co-spatial with the extended X-ray source, we discover a faint diffuse halo of emission. To highlight this diffuse emission, we created an image from the compact source-subtracted data at $30''$ resolution, which we display in \autoref{sec:app_radio}. 
We measured the flux densities of these sources at 54 and 144\,MHz and report them in \autoref{tab:fluxes}.
We also imaged the LoTSS data in circular polarization to check for the presence of a possible radio pulsar, no emission above $5\sigma_\mathrm{rms}$ ($500\,\mathrm{\mu{Jy}\,beam^{-1}}$) was found.

\begin{figure}
    \centering
    \includegraphics[width=0.99\linewidth]{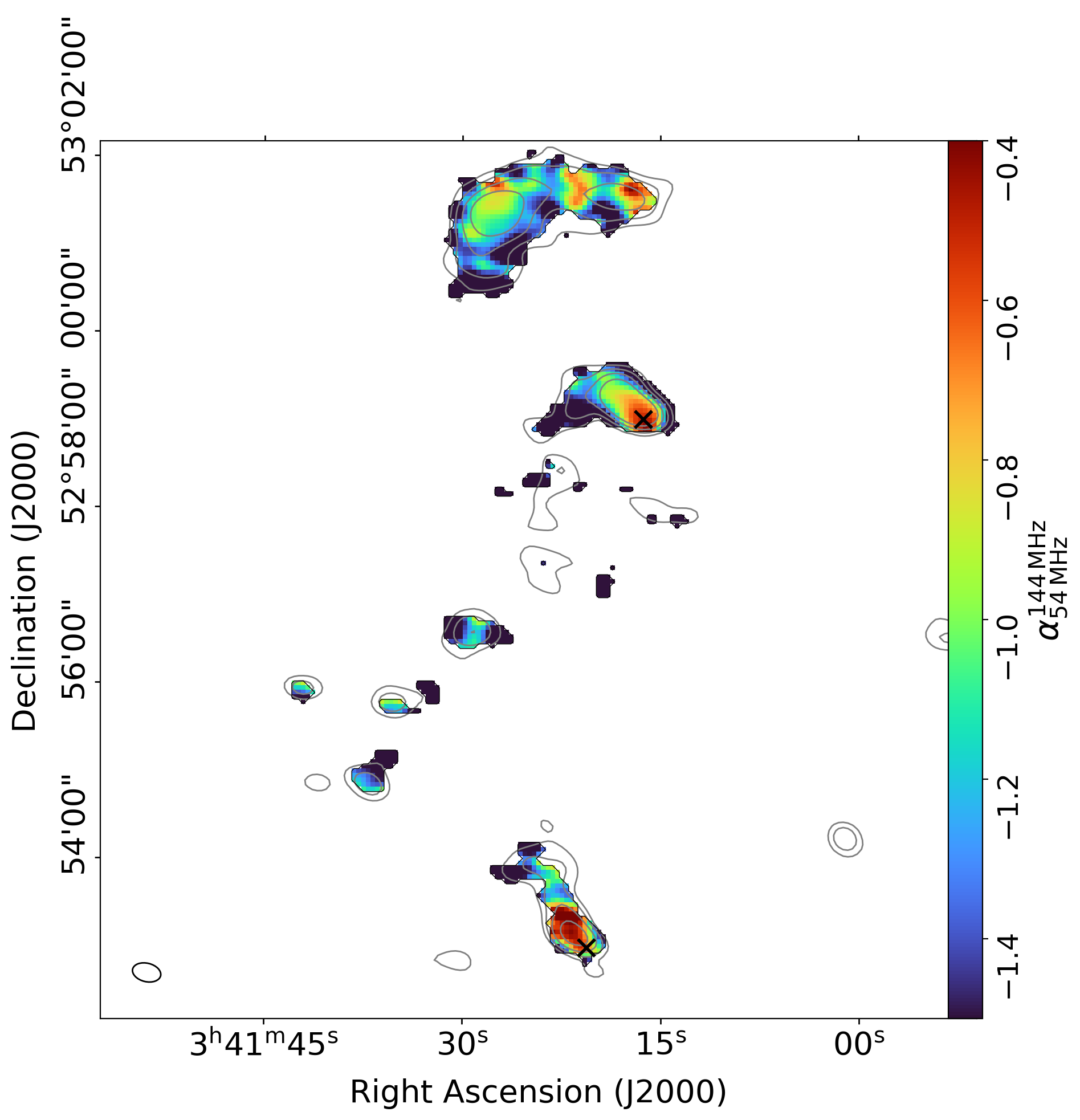}
    \caption{Radio spectral index map at a resolution of $20''\times13''$. Grey contours correspond to $1\,\mathrm{Jy\,beam^{-1}}\times[1,2,4,8]$ in the 144\,MHz map at the same resolution, black crosses mark the location of candidate IR host galaxies. }
    \label{fig:spidx}
\end{figure}

Based on the LBA and HBA images, we investigate the spectral properties of the radio sources. The arc-shaped source has an integrated spectral index of $\alpha=-1.18\pm0.15$, while for the tailed sources, we find values of $\alpha=-0.97\pm0.15$ and $\alpha=-0.93\pm0.17$ for the northern and southern source, respectively. For the diffuse halo emission co-spatial to the X-rays, we find an steep spectrum of $\alpha=-1.36\pm0.38$. We also created a spectral index map (displayed in \autoref{fig:spidx}) from the 54 and 144\,MHz images, as described in \autoref{sec:app_radio}.
While its resolution and quality is limited by the 54\,MHz data, we nevertheless find that the arc-like source generally features a flatter spectrum towards the north and a steeper spectrum towards the south, while the spectra of the two tailed sources are flat ($\alpha\sim -0.5$) towards the south and steep ($\alpha< -1$) towards the north. 
\subsection{UKIDSS GPS near-infrared galaxy density}\label{sec:nir}
We investigated the presence of possible NIR counterparts of the radio sources in the UKIDSS Galactic Plane Survey (UGS, \citealt{Lucas2008UKIDSSGalacticPlane}). While no prominent galaxies are found co-spatial to the candidate relic, both candidate TRGs show possible counterparts, marked as red circles in the insets in \autoref{fig:combined}. 

\begin{figure*}
\sidecaption
  \includegraphics[width=12cm]{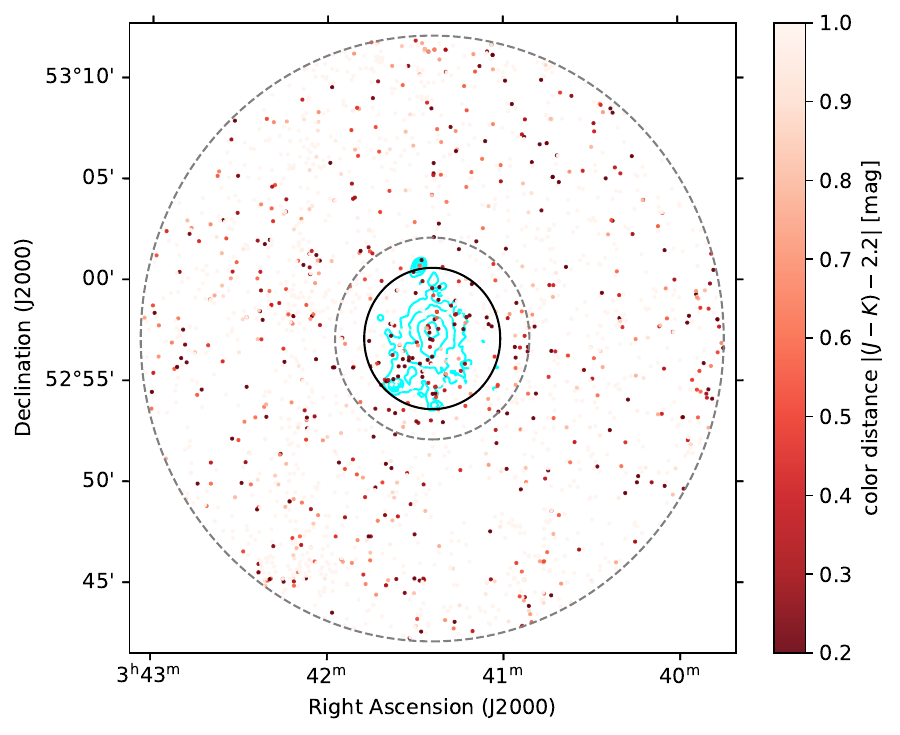}
    \caption{UKIDSS GPS galaxies, the marker color corresponds to the difference in $(J-K)$ color from a value of 2.2. The black circle marks the target and the dashed circular annulus the background region. In cyan, we show the X-ray contours.}
    \label{fig:coldist}
\end{figure*}

To investigate the presence of a galaxy over-density behind the Galactic plane, we employ the NIR ($J$,\,$H$,\,$K$ band) source catalog of the UGS, only considering sources detected in all three bands that are reported to have a probability $\geq95\%$ of being a galaxy. We correct the measurements for dust absorption following \cite{Schlafly2011MeasuringReddeningSloan}.
We compare the galaxy density in two different regions centered on the X-ray peak: one circular `target' region with a radius of $3.5'$, and a circular annulus `background' region with inner and outer radii of $5'$ and $15'$, respectively.
We estimate the expected number of galaxies in the target region $\lambda$ based on the target region area $A_\mathrm{t}$  multiplied by the galaxy density in the background region: $\lambda = {A_\mathrm{t}}N_\mathrm{bg}/{A_\mathrm{bg}}$, where $N_\mathrm{bg}$ and $A_\mathrm{bg}$ are the galaxy count and area of the background region. The uncertainty of this estimate is $\sigma_{\lambda}=\sqrt{\lambda}$, following the Poisson statistics.
This yields an expectation of $\lambda = 106.7\pm10.3$ galaxies in the target region, while it contains 164 objects. This corresponds to a $5.1\sigma$ over-density. Here, we neglected the uncertainty on the background estimate given its much larger area.
In the cluster scenario, the over-density should be strongest in early-type galaxies at a similar redshift, thus, with similar red colors.  
To test this, we repeated the statistical calculation, this time only considering galaxies with a $(J-K)$ color $2.2\pm0.5$. In this color interval, $\lambda = 15.3\pm3.9$ galaxies are expected, while 67 are contained within the target region. The significance of the over-density grows to $9.7\sigma$. 
In \autoref{fig:coldist}, we display the UGS galaxies and their ($J-K$) color on top of the regions used in this analysis. 
\section{Discussion}\label{sec:discussion}
In the following, we  will  discuss our newly detected radio sources in the context of the possible interpretation of XMMU J034124.2+525720 as a galaxy cluster. We classify the northern `Arc' radio source as a candidate radio relic based on its steep spectrum, its elongated morphology and its location and orientation with respect to the X-ray emission. The source is located in the direction of the major axis of the X-ray source, as expected for an ICM shock propagating along the merger axis.
In this scenario, shock acceleration would take place at the northern edge of the emission, where the spectrum should be flatter (typically $\alpha\sim-0.7$), whereas the downstream region should be steeper due to spectral aging ($\alpha<-1$). The quality of the spectral index map shown in \autoref{fig:spidx} is not sufficient to draw firm conclusions, but it is generally in agreement with such a scenario. 
In the interpretation as a radio relic, the bright area in the eastern part of the source would be a surface brightness fluctuation, for example due to a spatial variation in the seed electron density or magnetic field strength.
We note that the interpretation as a radio relic requires the source to be unrelated to the nearby X-ray point source.
We tentatively classify the low-surface brightness diffuse emission co-spatial to the X-ray source as a radio halo due to its similar morphology to the X-ray source and the steep spectral index ($\alpha\sim-1.4$). Both radio relic and radio halo sources are commonly found in massive merging galaxy clusters \citep{Cuciti2021}.
We further classify both the northern and southern tailed sources as candidate TRG based on their morphology, steep spectral indices ($\alpha\leq -0.9)$ and the superposition with bright red NIR galaxies. Their spectral index trend with flat ($\alpha\sim-0.5$) values toward the suggested host galaxy and steeper values along the tail is in agreement with this interpretation, as it resembles the characteristic spectral aging of cosmic rays along tailed radio galaxies \citep[][]{Edler2022}. TRGs are a tracer of over-dense regions such as galaxy clusters.

In \autoref{sec:xray}, we showed that the X-ray spectrum can be fitted as absorbed thermal ICM emission - this scenario is in fact slightly preferred to the non-thermal power-law fit. 
While the high extinction of the Galactic plane ($E(B-V)=1.7$\,mag) prevents us from identifying a possible cluster in the optical, the availability of the UGS data in the NIR, being substantially less affected by extinction, allowed us to study the distribution of galaxies. Indeed, we found a highly significant excess of galaxies with red $(J-K)$ colors. 
Given the combined evidence from the X-ray, radio and NIR data, and the lack of clear evidence for a pulsar, we favor the interpretation of XMMU\,J034124.2+525720 as a massive merging galaxy cluster.
Future observations in the hard X-ray regime will allow to obtain certainty about the nature of XMMU\,J034124.2+525720, as at these energies, the power-law and thermal models strongly deviate.

Similarly, the classification as a galaxy cluster would be certain by confirming the redshift.
The best-fitting value of $z\sim0.37$ from our APEC model is not perfectly constrained. 
A determination of the redshift is possible via NIR spectroscopy of the red galaxies in the vicinity of the X-ray source. Alternatively, deep high-resolution X-ray spectroscopy would allow for the detection of ICM emission lines at a high signal-to-noise ratio.
With the available data, estimates of the redshift can be obtained from the size of the radio sources and the brightness of the NIR galaxies. The vast majority of known relics have projected sizes $\geq500$\,kpc \citep{Jones2023PlanckClustersLOFAR}, translating to a redshift $z>0.2$. Assuming a typical size of the radio halo of 1\,Mpc broadly suggests $z\sim0.26$. The $K$-band magnitude of the candidate brightest cluster galaxy ($\alpha_\mathrm{J2000}{=}03^\mathrm{h}41^\mathrm{m}24^\mathrm{s}$, $\delta_\mathrm{J2000}{=}{+}52\degr57\arcmin05\arcsec$) is $K=14.2$\,mag. Assuming the galaxy's absolute magnitude is within a scatter of $\sim0.44\,$mag around $M_K=-26.12$ \citep{Lin2004}, we obtain a rough redshift range  $0.23 \le z \le 0.36 $ after applying $k$-correction  following \citet{Manucci2001}.

If the cluster scenario is correct and the X-ray source is not a PWN, the question of what object powers the UHE emission in its vicinity remains. 
In principle, UHE emission in clusters can originate from active galactic nuclei or from hadronic collisions of cosmic rays in the ICM. 
However, distant UHE sources are subject to strong absorption via $e^+e^-$-pair production with extragalactic photon fields. Furthermore, in their analysis of nearby galaxy clusters using LHAASO, \citet{cao25} did not detect even the nearest systems.
Thus, 1LHAASO~J0343+5254u is likely associated with a yet-to-be-discovered Galactic source. \cite{bangale23} noted that Fermi source 4FGL~J0340.4+5302 \citep{abdolahi20} is  possibly associated with the UHE emission; \cite{desarkar24} further explore the possibility that a pulsar TeV halo could be responsible for the gamma-ray emission. Finally, \cite{tsuji25} identified several molecular clouds that could serve as proton–proton collision targets, producing hadronic gamma-rays via neutral pion decay. 
Future multi-wavelength follow-up observations of this region may shed light on the origin of the UHE emission.
\section{Conclusion}\label{sec:conclusion}
We suggest an alternative interpretation of the extended X-ray source XMMU\,J034124.2+525720, initially reported by DK25 as a candidate PWN, as a galaxy cluster. We showed that a thermal plasma model with a redshift $z\approx0.37$ provides an equal, and even slightly preferred fit to the X-ray spectrum, and that the X-ray source is co-spatial with a $9.7\sigma$ strong over-density of red $(J-K) \sim 2.2$ NIR galaxies. We present novel LOFAR data and report the detection of sources which we classify as a (candidate) radio relic, radio halo and TRGs. Given the lack of clear evidence for the existence of a pulsar and the compelling evidence for the cluster scenario, we prefer the interpretation of XMMU\,J034124.2+525720 as a previously unrecognized merging cluster. The interpretation of the X-ray source as a galaxy cluster means that the counterpart of 1LHAASO J0343+5254u remains unknown. Follow-up observations in the hard X-rays will be critical to obtain clarity on the classification of the X-ray emission, while NIR spectroscopy of the galaxies or deep high-resolution spectroscopy of the X-ray source will be necessary to confirm the cluster's redshift.

\begin{acknowledgements}
The authors thank F. Gastadello for useful discussions.
This paper uses data obtained with the LOFAR telescope 
(projects LT14\_004 and
LT16\_004). 
LOFAR (van Haarlem et al. 2013) is the Low Frequency Array designed and constructed by
ASTRON. It has observing, data processing, and data storage facilities in several countries, which are owned by various parties (each with their own funding sources), and that are collectively operated by the ILT foundation under a joint scientific policy. The ILT resources have benefited from the following recent major funding sources: CNRS-INSU, Observatoire de Paris and Université d'Orléans, France; BMBF, MIWF-NRW, MPG, Germany; Science Foundation Ireland (SFI), Department of Business, Enterprise and Innovation (DBEI), Ireland; NWO, The Netherlands; The Science and Technology Facilities Council, UK; Ministry of Science and Higher Education, Poland; The Istituto Nazionale di Astrofisica (INAF), Italy. This research made use of the Dutch national e-infrastructure with support of the SURF Cooperative (e-infra 180169) and the LOFAR e-infra group. The Jülich LOFAR Long Term Archive and the German LOFAR network are both coordinated and operated by the Jülich Supercomputing Centre (JSC), and computing resources on the supercomputer JUWELS at JSC were provided by the Gauss Centre for Supercomputing e.V. (grant CHTB00) through the John von Neumann Institute for Computing (NIC).
This research made use of the University of Hertfordshire high-performance computing facility and the LOFAR-UK computing facility located at the University of Hertfordshire and supported by STFC [ST/P000096/1], and of the Italian LOFAR IT computing infrastructure supported and operated by INAF, and by the Physics Department of Turin university (under an agreement with Consorzio Interuniversitario per la Fisica Spaziale) at the C3S Supercomputing Centre, Italy. 
\end{acknowledgements}
\bibliographystyle{aa}
\bibliography{references}

%%%%%%%%%%%%%%%%%%%%%%%%%%%%%%%%%%%%%%%%%%%%%%%%%%%%%%%%%%%%%%%
% Appendices must be placed after   \end{thebibliography}
% They will be placed automatically on a new page.
%%%%%%%%%%%%%%%%%%%%%%%%%%%%%%%%%%%%%%%%%%%%%%%%%%%%%%%%%%%%%%%
\begin{appendix}
%%%%%%%%%%%%%%%%%%%%%%%%%%%%%%%%%%%%%%%%%%%%%%%%%%%%%%%%%%%%%%%
% In the PDF output, floats should be placed
% under their own appendix, not before the title, nor after the
% title of the next appendix.

% In short appendices, onecolumn floats (\figure*
% or \table*) will generate a blank page.
% To prevent this behaviour, a few examples are provided here. 

% In case you have a lot of floating objects for little text and the 
% LaTeX engine moves the floats away from their context, the command
% \FloatBarrier of the “placeins” package will empty the
% float buffer and place all stored floats in the continuity.

% If you still encounter problems with wide floats placement,
% just use the onecolumn environment throughout the appendices.
%%%%%%%%%%%%%%%%%%%%%%%%%%%%%%%%%%%%%%%%%%%%%%%%%%%%%%%%%%%%%%%

%____________________________________________________________
%       Wide floats at the start of an appendix: first method
%-------------------------------------------------------------
% To prevent a blank page after the start of an appendix:
% - Switch to one \onecolumn first
% - Declare the section title
% - Declare the onecolumn float with the parameter [h!]
% - Revert to \twocolumn at the end of the section
%\onecolumn
\section{LOFAR data analysis}\label{sec:app_radio}
\FloatBarrier
The coordinates of the X-ray source are covered by the pointings P053+54, P055+51 and P057+54 of the common LoTSS and LoLSS survey grid. In both cases, we carry out the source extraction and re-calibration procedure originally described in \citet{vanWeeren2021LOFARObservationsGalaxy}. This means that from each pointing at both frequencies, we subtract all sources in the field of view from the visibilities except for a circular region of $24'$ radius centered on the X-ray source. 
Then, we phase shift all observations to this center and correct the visibilities for the primary beam effect at the new phase center. Then, we carry out a self-calibration of the data. For the for LBA and HBA observations, we solve for a `tecandphase' constraint on 16\,s of time resolution. For the HBA data, we additionally solve for a `scalarcomplexgain' constraint on 30\,min time resolution. The final images are created with \texttt{WSClean} \citep{Offringa2014WSCLEANImplementationFast} and corrected for the primary beam attenuation during imaging. 
The image noise level is $\sigma_\mathrm{rms}=100\,\mathrm{\mu{Jy}\,beam^{-1}}$ for the HBA and $\sigma_\mathrm{rms}=2.1\,\mathrm{mJy\,beam^{-1}}$ for the LBA, at angular resolutions of $5''\times8''$ and  $13''\times19''$, respectively.

\begin{figure}
    \centering
    \includegraphics[width=0.99\linewidth]{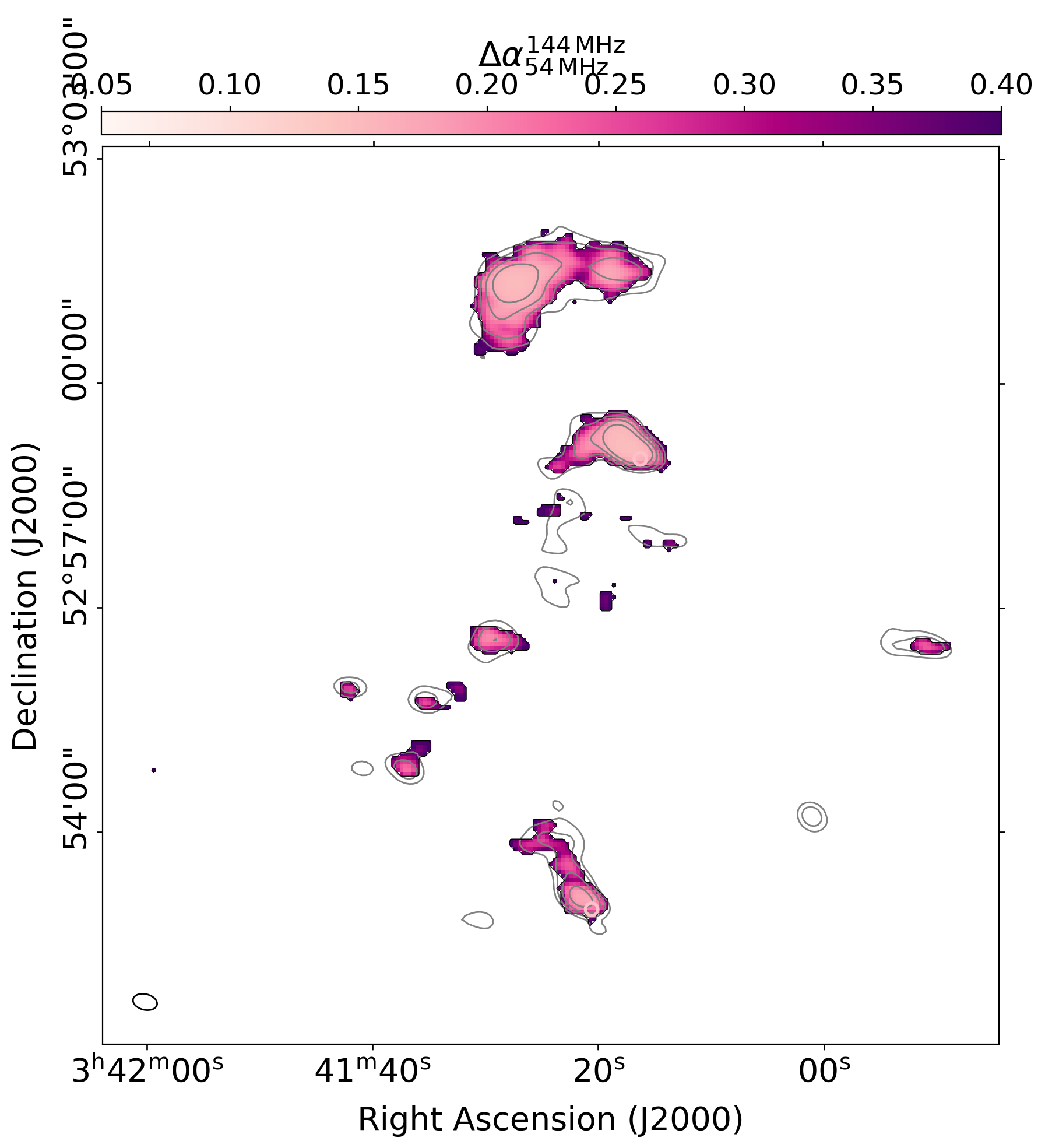}
    \caption{LOFAR spectral index uncertainty map at a resolution of $19.7''\times12.6''$.}
    \label{fig:sierr}
\end{figure}

From these, we create a spectral index map by regridding the images to the same pixel grid and convolving the HBA map to the synthesized beam shape of the LBA map. For all pixels above three times the root-mean-square (rms) noise, we calculated the spectral index according to:
\begin{equation}
    \alpha_{54}^{144} = \frac{\log({S_{54}/S_{144}})}{\log({54/144})}.
\end{equation} 
The spectral index map is shown in \autoref{fig:spidx} and the the corresponding uncertainty map is shown in \autoref{fig:sierr}.

\begin{figure}
    \centering
    \includegraphics[width=0.99\linewidth]{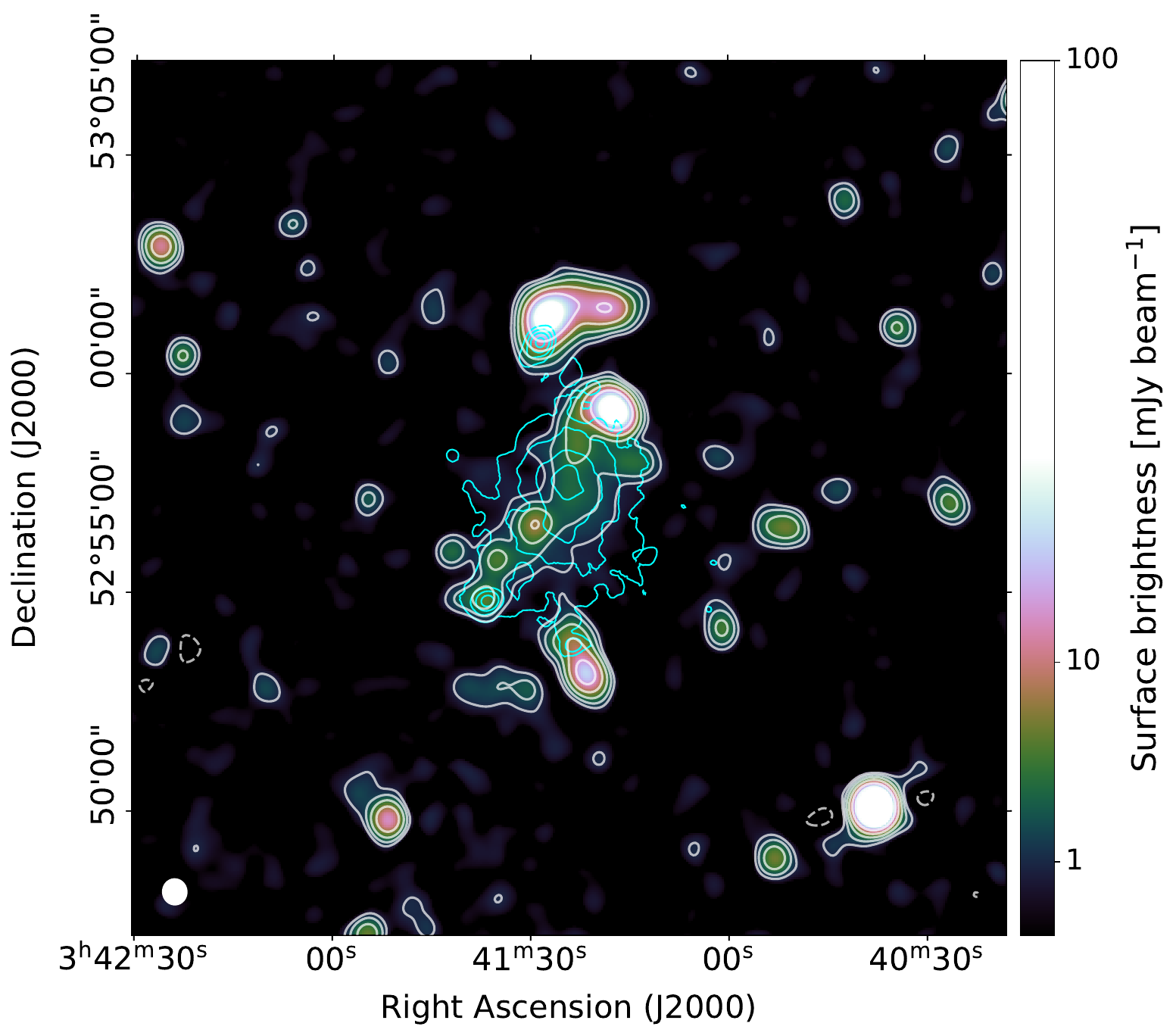}
    \caption{Compact-source subtracted radio map at 144\,MHz, the noise level is $\sigma_\mathrm{rms}=299\,\mathrm{\mu{Jy}\,beam^{-1}}$ at $33''\times35''$ resolution. The contours start at $3\sigma_\mathrm{rms}$ and increase in powers of two. In cyan, we show the $0.5-7$\,keV X-ray contours, starting at $5\times10^{-6}\,\mathrm{cts\,s^{-1}\,arcsec^{-2}}$ and increasing in factors of two.}
    \label{fig:hbalres}
\end{figure}
\FloatBarrier
\section{X-ray re-analysis}

\begin{figure}
    \centering
    \includegraphics[width=0.75\linewidth,angle=-90]{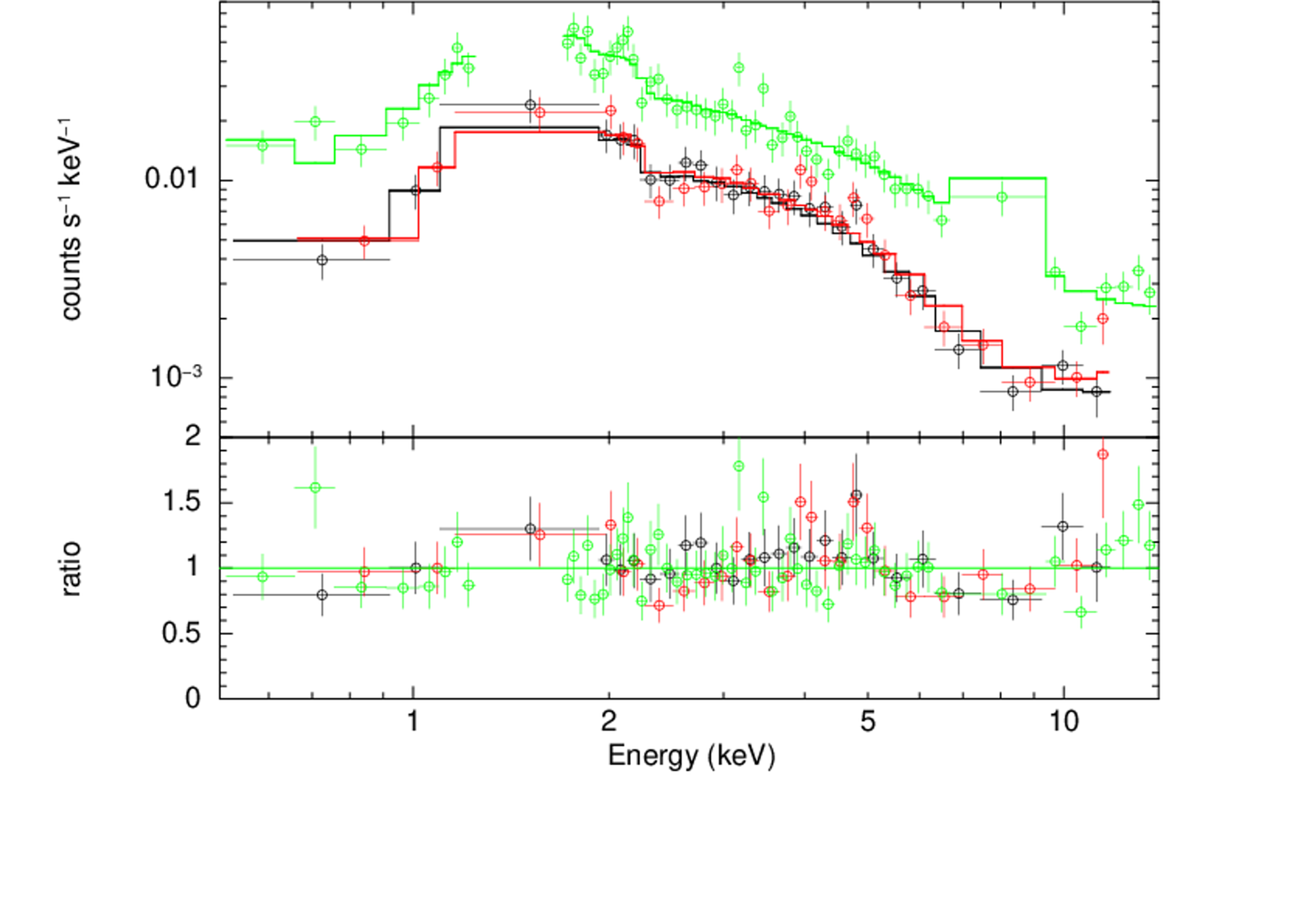}
    \vspace{-8mm}
    \caption{Observed source spectrum (black=MOS1, red=MOS2, green=pn) with power-law best-fit model reported. The spectra of the three cameras and three observations were jointly fitted, but the results of a single ObsID (0923400401) were reported for clarity. For energy ranges 1.2-1.9 keV (for the MOS detectors) and 1.2-1.7 keV and 7-9.2 keV (for the pn detector) were excluded during the fit. Each spectral bin was rebinned to a minimum significance of 5$\sigma$ for visualization purposes.}
    \label{fig:xrayspec_power}
\end{figure}

\end{appendix}
\end{document}